\documentclass[11pt]{article}
\usepackage{moriond,epsfig}

\def\be{\begin{equation}}
\def\ee{\end{equation}}
\def\bea{\begin{eqnarray}}
\def\eea{\end{eqnarray}}

\newcommand   {\e}  {{{e}}}
\newcommand   {\p}  {{{p}}}

\newcommand   {\Z}  {{{Z}}}

\renewcommand {\u}  {{{u}}}
\renewcommand {\d}  {{{d}}}

\begin{document}
\vspace*{4cm}
\title{ELECTRON-PROTON SCATTERING AT HIGH $Q^{2}$\\[.1cm]
Recent Results and Future Perspectives of Testing QCD \\
and Electroweak Theory at HERA}

\author{ H.-C. SCHULTZ-COULON }

\address{Universit\"at Dortmund, Experimentelle Physik V,\\
Otto-Hahn-Str.\ 4, Dortmund, Germany}

  \maketitle
  
  \abstracts{ The $\e^{+}\p$ and $\e^{-}\p$ scattering data 
  recorded at HERA during the recent years offer the possibility to 
  study electroweak effects in $\e\p$ interactions apparent at high 
  momentum transfers, $Q^{2}$, and 
  to reveal information on the proton parton densities at large values 
  of the Bjorken scaling variable~$x$.  From the neutral current cross 
  section measurements, H1 and ZEUS extract the generalized 
  structure function $x{\cal F}_{3}$, which 
  can be related to the valence quark content of the proton.  
  Individual quark densities are extracted by a global fit to the H1 
  neutral and charged current $\e^{+}\p$ and $\e^{-}\p$ data.  The new 
  results show the sensitivity of high $Q^{2}$ $\e\p$ data to the 
  structure of the proton and indicate what to expect from a 
  1~fb$^{-1}$ data sample to be taken by H1 and ZEUS until 2006 at the 
  upgraded HERA collider.  Future perspectives concerning the 
  investigation of electroweak effects and their utilization to 
  extract the parton content of the proton are shortly discussed.}

\section{Introduction}

At HERA 27.5~GeV positrons collide head on with 920~GeV protons 
\footnote{The proton beam energy has been increased from 820~GeV to 
920~GeV after the 1997 data taking period; data recorded before 1998 
are taken at a center-of-mass energy of 300~GeV.}, leading to a 
center-of-mass energy $\sqrt{s}$ of approximately 318~GeV. The HERA 
facility, with its two collider experiments H1 and ZEUS, therefore 
offers the unique possibility to probe the structure of the proton 
down to very small distances ($<10^{-18}\;$m) via $t$-channel exchange 
of highly virtual gauge bosons.

\begin{figure}[tb]
\begin{minipage}{0.496\textwidth}
\centerline{\epsfig{figure=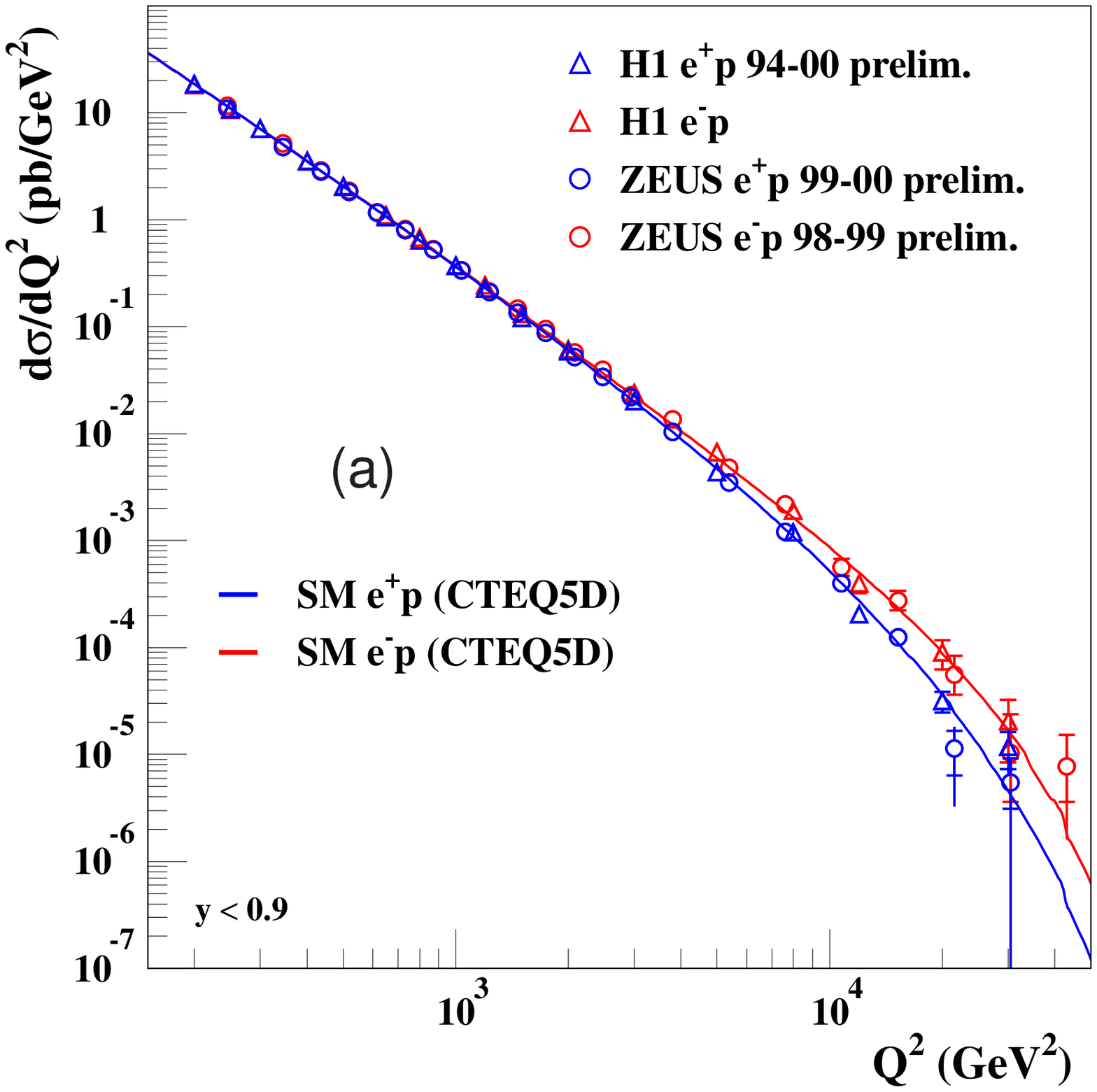,width=\textwidth}}
\end{minipage}
\begin{minipage}{0.496\textwidth}
\centerline{\epsfig{figure=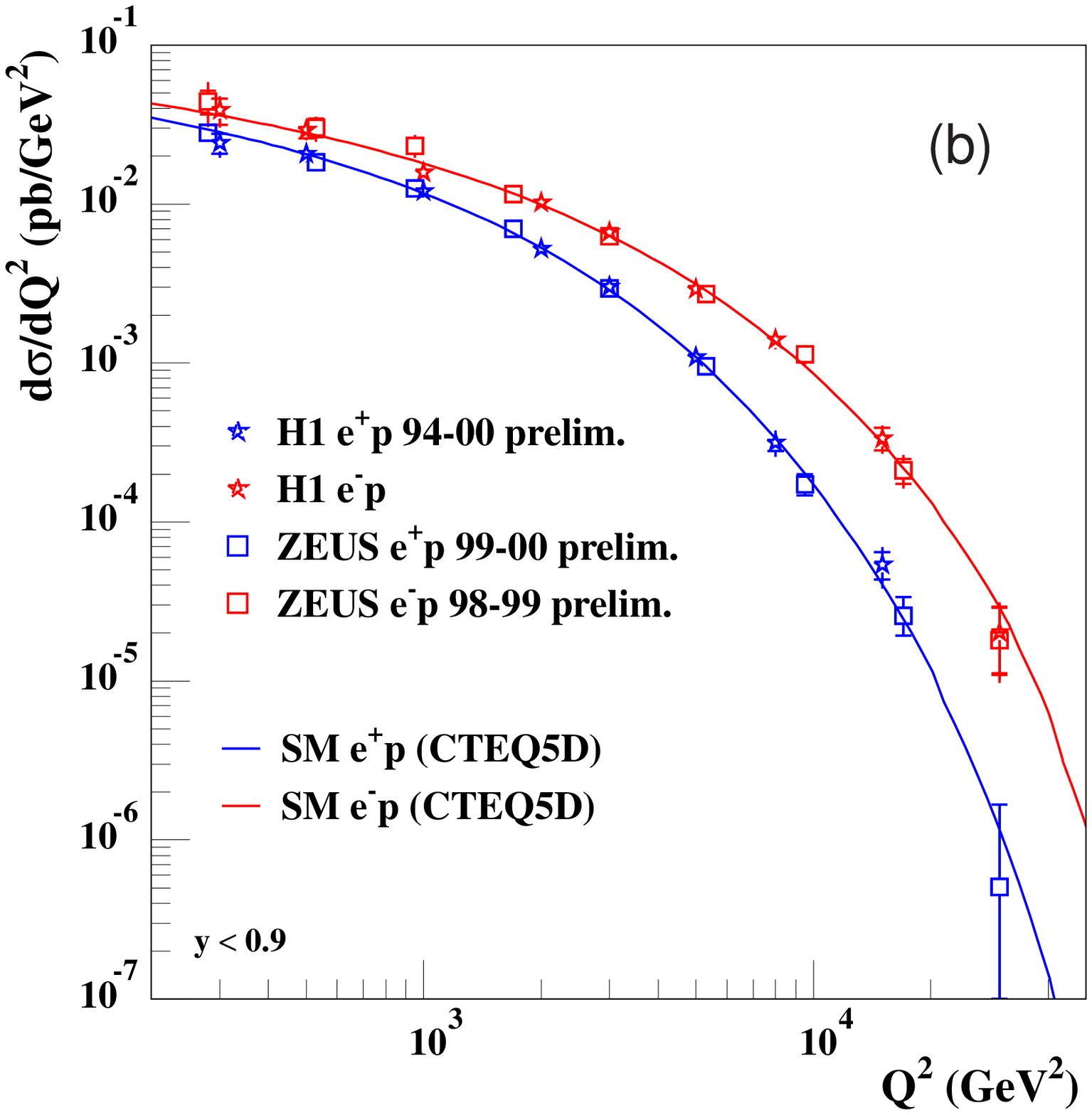,width=\textwidth}}
\end{minipage} 
\caption{The $Q^{2}$ dependence of the NC (a) and CC (b) cross 
sections, $\d\sigma/\d Q^{2}$, for $\e^{+}\p$ and $\e^{-}\p$ data and the 
corresponding ${\cal SM}$ expectation evaluated using the 
CTEQ5D~\protect\cite{Lai:1999wy} parton densities.  The statistical 
uncertainties are indicated by the inner error bars, while the full 
error bars show the total statistical and systematic uncertainties 
added in quadrature.}
\label{fig=NCandCC}
\end{figure}

The dominant process for deep-inelastic $\e\p$ scattering (DIS) is 
given by the exchange of a photon between the incoming lepton and a 
quark of the proton.  However, at high momentum transfers, $Q^{2}$, 
the relative contribution from neutral (NC) and charged current (CC) 
reactions with exchange of a massive vector boson, a $Z^{0}$ or a 
$W^{\pm}$, becomes important, allowing the investigation of 
electroweak effects in lepton-proton reactions.  With the statistics 
collected at HERA until the end of 2000 precise $\e^{+}\p$ and 
$\e^{-}\p$ cross section measurements~\cite{Adloff:1999ah}$^{\rm 
-}\,$\cite{EPS2001:2001nm}
have become available, which at high $Q^{2}$ clearly reveal the 
expected dependence on the lepton beam charge predicted by the 
Standard Model (${\cal SM}$), e.g. an increased NC cross section for 
$\e^{-}\p$ with respect to $\e^{+}\p$ scattering due to positive 
interference between the $\gamma$- and $\Z$-exchange.  This can be 
seen from figure~\ref{fig=NCandCC}, which compiles the most recent H1 
and ZEUS results for the differential neutral and charged current 
cross sections, $\d\sigma/\d Q^{2}$, reaching values of $Q^{2}$ up to 
$30000$~GeV$^{2}$.

Exploitation of the observed dependence on the lepton beam charge 
yield access to information on the valence quark content of the proton 
at high values of the Bjorken scaling variable~$x$, as will be 
discussed in the following sections.  The data do, however, not yet 
allow for a precise extraction of electroweak parameters.  This will 
only be possible after the HERA~2 run has been completed, for which a 
large increase in statistics is anticipated.  Some of the prospects 
concerning electroweak physics and the determination of parton 
densities at high~$Q^{2}$ and high~$x$ are considered in 
section~\ref{sec=Future Perspectives}.

\section{Neutral Current Reactions}
\label{sec=NC}

\begin{figure}[tb]
\label{fig=NCredxsec} 
\begin{minipage}{0.496\textwidth}    
\centerline{\epsfig{figure=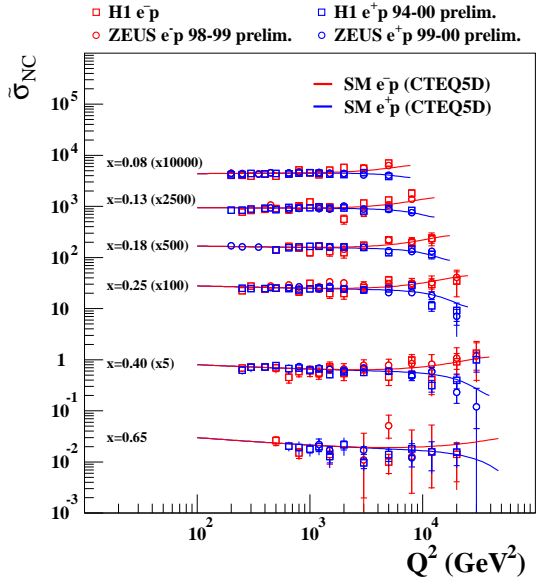,width=\textwidth}}
\end{minipage}
\begin{minipage}{0.476\textwidth} 
\vspace*{3.0cm} \caption{The NC reduced cross section 
$\tilde{\sigma}_{\rm NC}(x,Q^{2})$ at high~$x$ compared to the ${\cal 
SM}$ prediction evaluated using the CTEQ5D~\protect\cite{Lai:1999wy} 
parton densities.  The inner error bars represent the statistical 
error, and the outer error bars the total error.}
\end{minipage}
\end{figure}

The Born cross section~\cite{Derman:sp,Ingelman:1987zv} for the NC DIS reaction 
$\e^{\pm}\p \rightarrow \e^{\pm}X$
is given by
\begin{equation}
\label{eqn=NCgen}
     \frac{\d^2\sigma^{\pm}_{\rm NC}}{\d x\,\d Q^2} =  
           \frac{2\pi\alpha^2}{xQ^4} 
               \left\{ Y_+(y)\,{\cal F}_2(x,Q^2) \mp
                       Y_-(y)\,x{\cal F}_3(x,Q^2) - y^{2}{\cal 
		       F}_{L}(x,Q^2) \right\}
\end{equation}
where $\alpha$ is the fine structure constant and the functions 
\mbox{$Y_\pm=1\pm(1-y)^2$} describe the helicity dependence of the 
electroweak interactions.  The longitudinal structure function ${\cal 
F}_L$ contributes only at high~$y$ (i.e.\ small~$x$ and small~$Q^{2}$ 
for HERA kinematics) and is neglected in the following discussion 
\footnote{In general both experiments estimate the ${\cal F}_{L}$ 
contributions from QCD fits and take them into account when deriving 
results at very high~$y$ or calculating systematic errors.  A recent 
H1 determination~\cite{EPS2001:2001im} of ${\cal F}_L$ at high $Q^{2}$ 
shows good agreement between the QCD predictions and the measurement 
justifying this approach.}.  The partonic structure of the proton 
is then contained in the generalized structure functions ${\cal F}_2$ and 
${\cal F}_3$.
For unpolarized beams they can be written as
\begin{equation}
\label{eqn=F2def}
  \left(
    \begin{array}{r} 
      {\cal F}_2(x,Q^2) \\ 
      x {\cal F}_3(x,Q^2) 
    \end{array}
  \right) 
  = \sum_{q={\rm quarks}} \!\! x
  \left(
    \begin{array}{c}
      C^q_2(Q^2)\,\left[q(x,Q^2) + \bar{q}(x,Q^2)\right] \\
      C^q_3(Q^2)\,\left[q(x,Q^2) - \bar{q}(x,Q^2)\right]
    \end{array}
  \right) \,\raisebox{-4mm}[-4mm]{.}
\end{equation}
Here, $q$ and $\bar{q}$ are the quark densities depending on $x$ and $Q^2$
alone, while the coefficient functions $C^q_2$ and $C^q_3$ can be expressed
in terms of precisely measured electroweak parameters. In lowest order 
they are given by
\begin{eqnarray}
     C^{q}_{2} & = & e_{q}^{2} - 2 e_{q} v_{e} v_{q} \chi_{Z} + (v_{e}^{2} + 
     a_{e}^{2})(v_{q}^{2} + a_{q}^{2}) \chi_{Z}^{2} \\
     C^{q}_{3} & = & - 2 e_{q} a_{e} a_{q} \chi_{Z} + 
     4v_{e}a_{e}v_{q}a_{q} \chi_{Z}^{2} \;\; , \label{eqn=eweak}
\end{eqnarray}
where $v_{e}$, $v_{q}$ and $a_{e}$, $a_{q}$ are the vector and axial 
couplings of the $\Z$-boson to quarks and electrons, $e_{q}$ 
represents the quark charge and $\chi_{Z} = \kappa_{w} 
Q^{2}/(Q^{2}+M_{\Z}^{2})$ is the propagator term with $\kappa_{w}^{-1} 
= 4 \sin^{2} \theta_{W} \cos^{2} \theta_{W}$.  Generally it is 
convenient to derive the neutral current reduced cross section in 
which the dominant part of the $Q^{2}$ dependence of $\d^{2}\sigma/\d 
x \d Q^{2}$ due to the $\gamma$-propagator is removed
\begin{equation}
    \tilde{\sigma}_{\rm NC} =  \frac{1}{Y_{+}} 
    \frac{Q^{4}x}{2\pi\alpha^{2}} \frac{\d^{2}\sigma_{\rm 
    NC}}{\d x \d Q^{2}} \;\; \raisebox{-2.5mm}[-2.5mm]{.}
\end{equation}    
In this form the differences between $\e^{-}\p$ and $\e^{+}\p$ 
scattering at high $Q^{2}$ are more clearly observed, as can be seen 
from figure~\ref{fig=NCredxsec}.  Up to $Q^{2}$ values of about 
1000~GeV$^{2}$ the $\e^{-}\p$ data are found to be in agreement with 
the $\e^{+}\p$ measurements as is expected for a solely electromagnetic 
interaction.  At larger values of $Q^{2}$, however, the $\e^{-}\p$ 
data lie generally above the $\e^{+}\p$ results.  This is compatible 
with a positive (negative) contribution from $x{\cal F}_{3}$ to the 
$\e^{-}\p$ ($\e^{+}\p$) cross section as given in 
equation~(\ref{eqn=NCgen}) and thus predicted by the Standard Model.

\begin{figure}[tb]
\hspace*{-.5cm}
\begin{minipage}{0.496\textwidth}
\centerline{\epsfig{figure=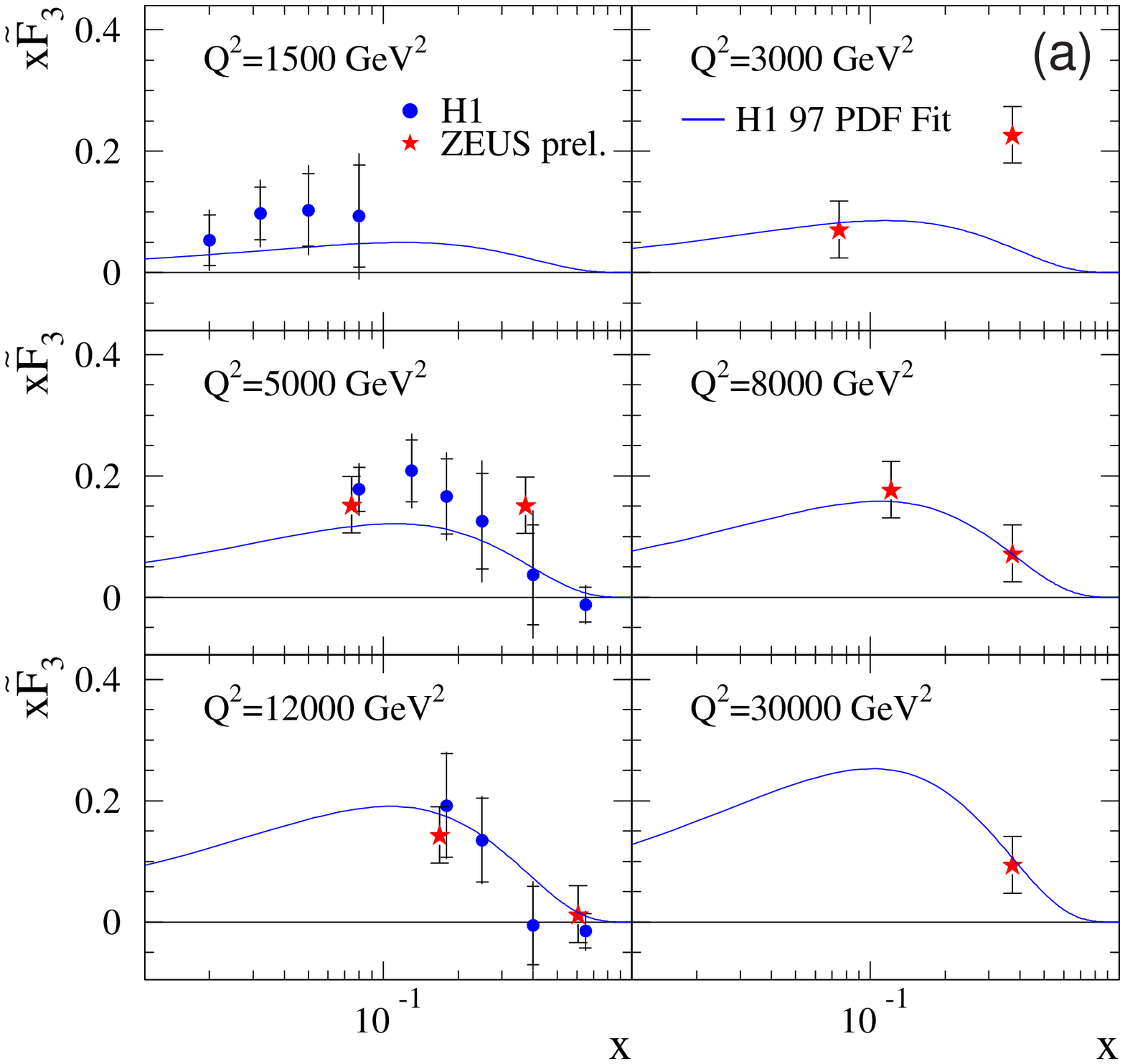,width=\textwidth}}
\end{minipage}\hspace*{.25cm}
\begin{minipage}{0.496\textwidth}
\vspace*{-.07cm}
\centerline{\epsfig{figure=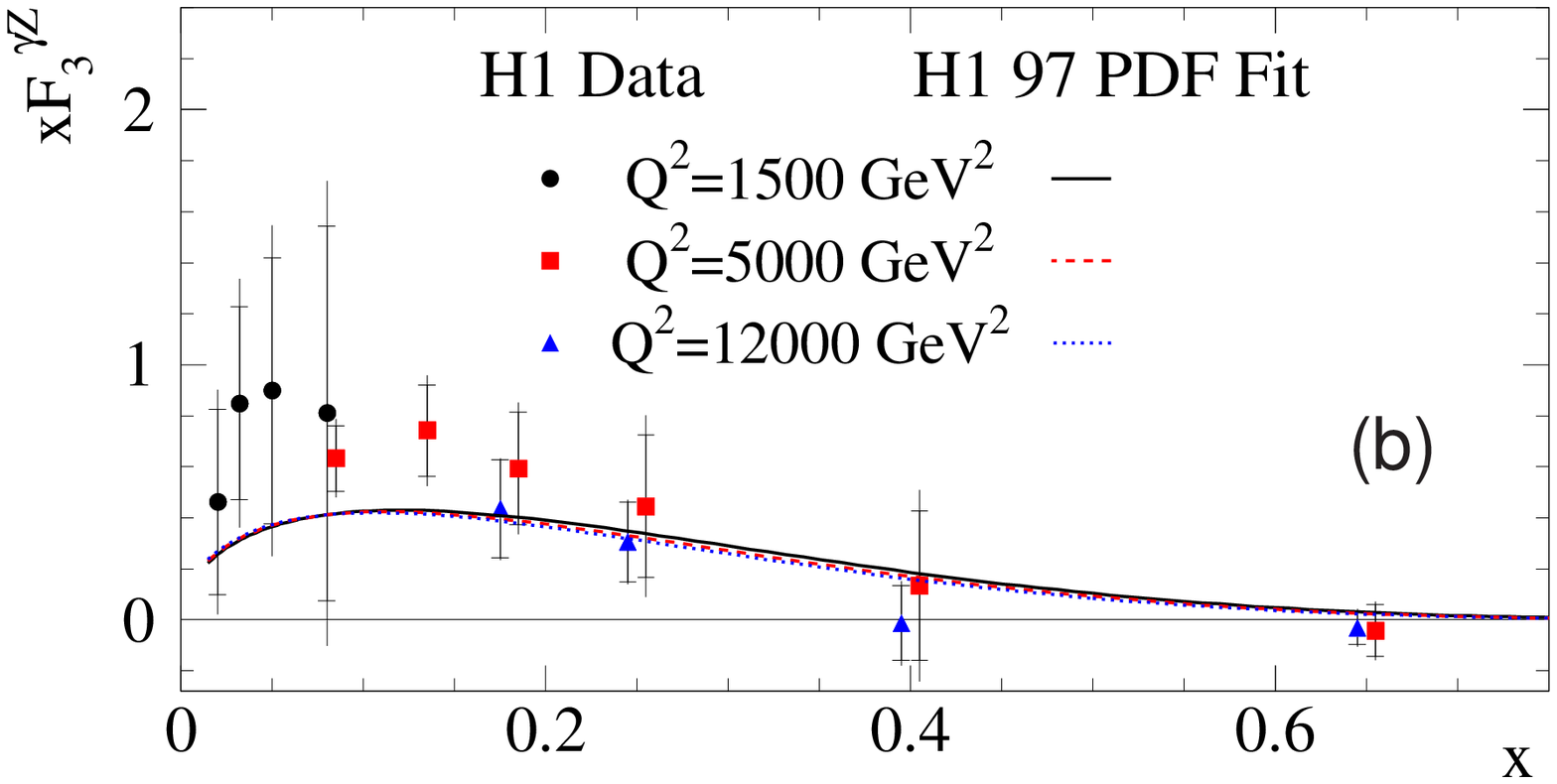,width=\textwidth}}
\vspace*{-.2cm} \caption{\label{fig=xF3} (a) The generalized structure 
function $x{\cal F}_{3}$, as extracted by H1 and ZEUS. The data are 
plotted at fixed $Q^{2}$ as a function of $x$.  (b) Structure function 
$xF_{3}^{\gamma\Z}$ as a function of $x$ for three different values of 
$Q^{2}$.  All data are compared to predictions from the H1 97 PDF 
fit~\protect\cite{Adloff:1999ah}.  Inner error bars represent the 
statistical, outer error bars the total error.}
\end{minipage}
\end{figure}

The difference between the $\e^{-}\p$ and $\e^{+}\p$ NC cross section 
can be used to extract the generalized structure function $x{\cal 
F}_{3}$ and --- since at HERA the dominant contribution to $x{\cal 
F}_{3}$ comes from the $\gamma\Z$--interference --- to evaluate the 
structure function $xF_{3}^{\gamma\Z}$, which is more closely related 
to the quark structure of the proton.  Figure~\ref{fig=xF3}a shows the 
$x{\cal F}_{3}$ $x$-dependence as measured by the H1 and ZEUS 
experiments in six different bins of $Q^{2}$.  As for the presented 
analyses the $\e^{-}\p$ data have been recorded at a different 
center-of-mass energy $\sqrt{s}=318$~GeV than the $\e^{+}\p$ data 
taken at $\sqrt{s}=300$~GeV, $x{\cal F}_{3}$ is evaluated using the 
following expression
\begin{equation}
    \tilde{\sigma}^{-}_{\rm NC} - \tilde{\sigma}^{+}_{\rm NC} = 
    x{\cal F}_{3}\left[\frac{Y_{-\;318}}{Y_{+\;318}} + 
    \frac{Y_{-\;300}}{Y_{+\;300}} \right] - {\cal F}_{L} \left[ 
    \frac{y^{2}_{318}}{Y_{+\;318}} -  \frac{y^{2}_{300}}{Y_{+\;300}} 
    \right] \;\;,
\end{equation}
where the contribution of ${\cal F}_{L}$ can again be neglected.  Here, 
$y_{318}$ and $y_{300}$ are the inelasticities, and $Y_{\pm\;318}$ and 
$Y_{\pm\;300}$ are the helicity functions as defined above, both 
evaluated for fixed $x$ and $Q^{2}$ and the corresponding 
center-of-mass energies $\sqrt{s}=318$~GeV and $300$~GeV. The results 
are in good agreement with the QCD prediction.

From these data H1 extracts~\cite{Adloff:2000qj} the structure 
function $xF_{3}^{\gamma\Z} = 2 e_{q} a_{q} \left[q-\bar{q}\right]$ 
dividing $x{\cal F}_{3}$ by the factor $-a_{e}\kappa_{w}Q^{2}/(Q^{2} + 
M_{\Z}^{2})$; the remaining contribution of order ${\cal O}(\chi_{Z})$ 
arising from pure $\Z$-exchange is estimated to be less than 3\% and 
hence neglected.  Figure~\ref{fig=xF3}b shows $xF_{3}^{\gamma\Z}$ as a 
function of $x$ for three values of $Q^{2}$.  The change of 
$xF_{3}^{\gamma\Z}$ at fixed $x$ over the measured $Q^{2}$ range is 
expected to be very small, as it arises only from QCD scaling 
violations for a non-singlet structure function.  In view of the large 
experimental errors it is thus possible to directly compare the 
results at the different $Q^{2}$ values.

The presented measurement yields first information on the valence 
quark content of the proton at high $Q^{2}$.  It is consistent with 
zero at large $x$ rising to a maximum at $x \approx 0.1$.  To quantify 
the level of agreement between data and theory the following sum rule 
has been formulated~\cite{Rizvi:2000qf} in analogy to the Gross 
Llewellyn-Smith sum rule~\cite{Gross:1969jf}: $\int_{0}^{1} 
F_{3}^{\gamma\Z} \d x \approx 5/3$.  Averaging the H1 data for 
different $Q^{2}$ at fixed $x$ by taking weighted means, integration 
yields~$\int_{.02}^{.65} = 1.88\pm 0.44$.  The corresponding integral 
obtained for the H1 97 PDF fit gives 1.11 and is thus found to agree 
within two standard deviations.
    
\section{Charged Current Reactions}
\label{sec=CC}

\begin{figure}[tb]
\vspace*{-.5cm}
\begin{minipage}{0.496\textwidth}
\centerline{\epsfig{figure=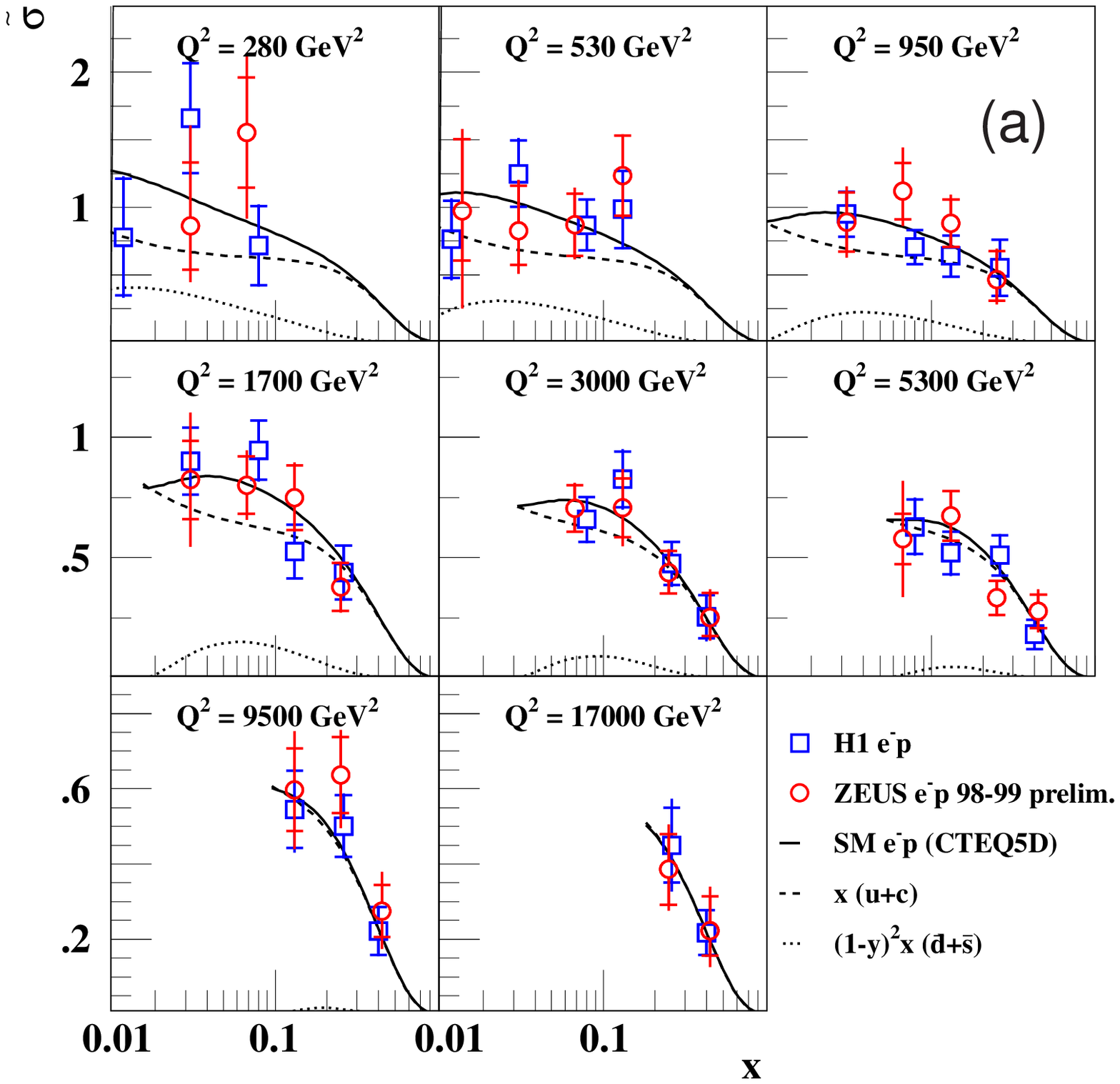,width=\textwidth}}
\end{minipage}
\begin{minipage}{0.496\textwidth}
\centerline{\epsfig{figure=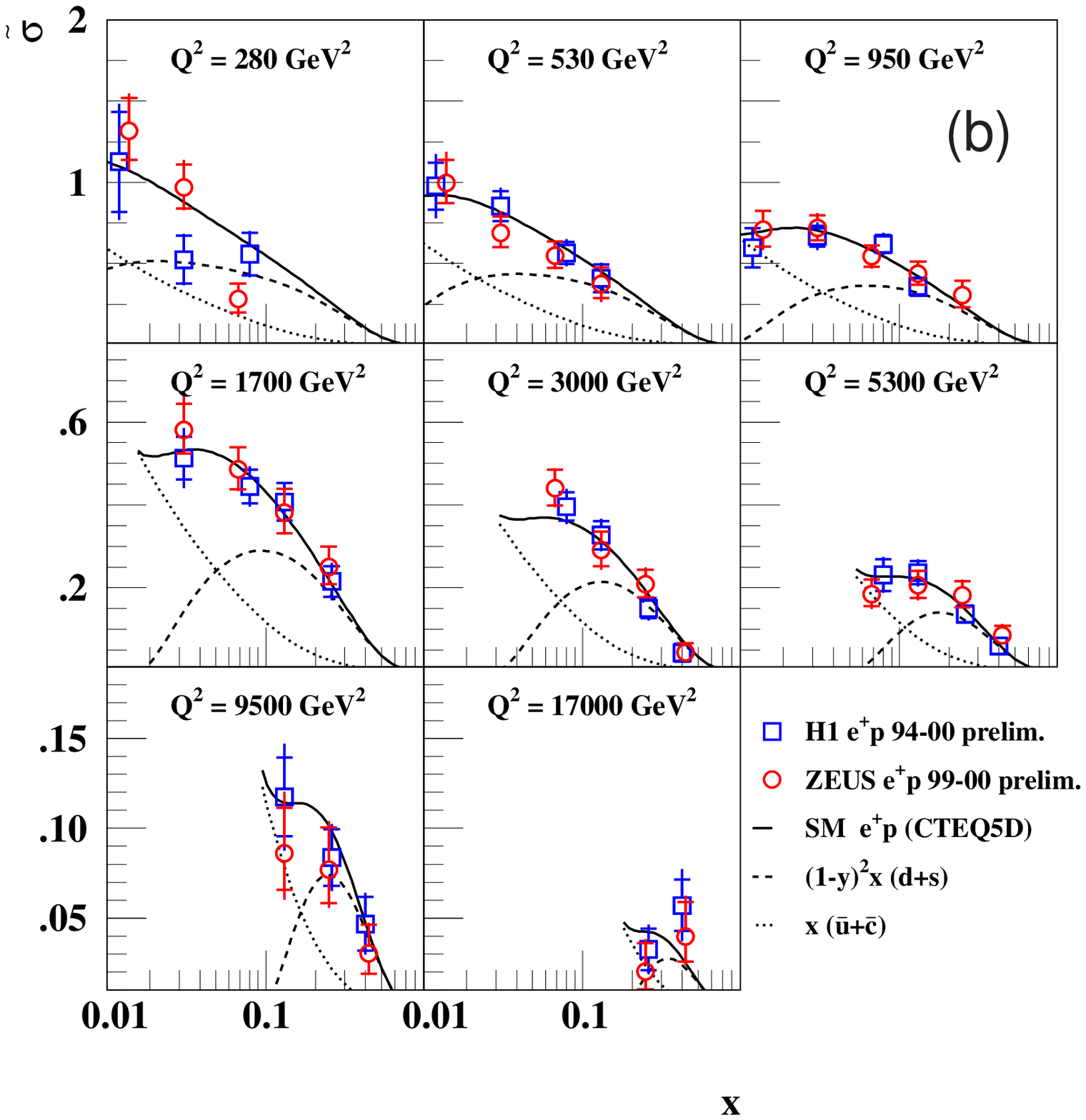,width=\textwidth}}
\end{minipage}
\caption{\label{fig=CCredxec}Charged current reduced cross section 
$\tilde{\sigma}_{\rm CC}(x,Q^{2})$ for $\e^{-}\p$ (a) and $\e^{+}\p$ 
(b) scattering as measured by the H1 and ZEUS collaborations.  The 
inner error bars represent the statistical error, and the outer error 
bars the total error of the measurement.  The data are compared to the 
${\cal SM}$ expectation evaluated using the 
CTEQ5D~\protect\cite{Lai:1999wy} parton densities.  The full lines 
represent the total cross section predictions; individual 
contributions from different quark flavours are also shown (dashed 
and dotted lines) to indicate the sensitivity of the measurement to the valence 
quark content of the proton at high~$x$.}
\end{figure}

In contrast to neutral current interactions, for which all quark and 
anti-quark flavours contribute, charged current $\e^{-}\p$ ($e^{+}\p$) 
reactions probe, in leading order, only down-type (up-type) quarks and 
up-type (down-type) anti-quarks, as they are mediated by the exchange 
of the $W^{-}$ ($W^{+}$) boson.  Charged current reactions thus allow 
for flavour-specific investigations of the parton momentum 
distributions and can provide additional information on the quark 
content of the proton at high~$x$ and high~$Q^{2}$.  
Since only the weak interaction contributes, the expression for the 
single-differential charged current cross section at the
Born-level~\cite{Ingelman:1987zv} can be written in a somewhat simpler 
form than in the neutral current case:
\begin{equation}
     \frac{d^2\sigma^{\pm}_{\rm CC}}{dx\,dQ^2} =  
           \frac{G_{F}^{2} M_{W}^{4}}{2\pi x} \frac{1}{(Q^{2} + 
           M_{W}^{2})^{2}} \tilde{\sigma}^{\pm}_{\rm CC} 
           \;\;\raisebox{-1mm}[-1mm]{,}
\end{equation}
with the reduced cross section given by
\begin{eqnarray}
    \tilde{\sigma}^{-}_{\rm CC} & = & x
       \left[ (u(x,Q^{2}) + c(x,Q^{2})) + (1-y)^{2} (\bar{d}(x,Q^{2}) + 
       \bar{s}(x,Q^{2})) \right] \\
    \tilde{\sigma}^{+}_{\rm CC} & = & x 
       \left[ (\bar{u}(x,Q^{2}) + \bar{c}(x,Q^{2})) + (1-y)^{2} 
       (d(x,Q^{2}) + s(x,Q^{2})) \right] \;\; \raisebox{-1.5mm}[-1.5mm]{.}
\end{eqnarray}
Here $u$, $c$, $d$ and $s$ are the quark and $\bar{u}$, $\bar{c}$, 
$\bar{d}$ and $\bar{s}$ the anti-quark distributions.  From these 
equations the sensitivity of CC reactions to the different quark 
densities becomes evident; at high~$x$, where the contribution from the 
sea can be neglected, $\e^{-}\p$ scattering provides direct access to 
the $u$-quark distributions, while positron-proton collisions 
probe the $d$-quark content of the proton.  This can be seen from 
figure~\ref{fig=CCredxec} showing the reduced $\e^{-}\p$ and 
$\e^{+}\p$ CC cross section, as measured by 
H1~\cite{Adloff:2000qj,EPS2001:2001im} and 
ZEUS~\cite{ICHEP2000:2000cm,EPS2001:2001cp}, together with the 
different contributions from quarks and anti-quarks according to the 
CTEQ5D~\cite{Lai:1999wy} parton densities.

\section{Parton Densities}

\begin{figure}[tb]
\begin{minipage}{0.696\textwidth}
\centerline{\epsfig{figure=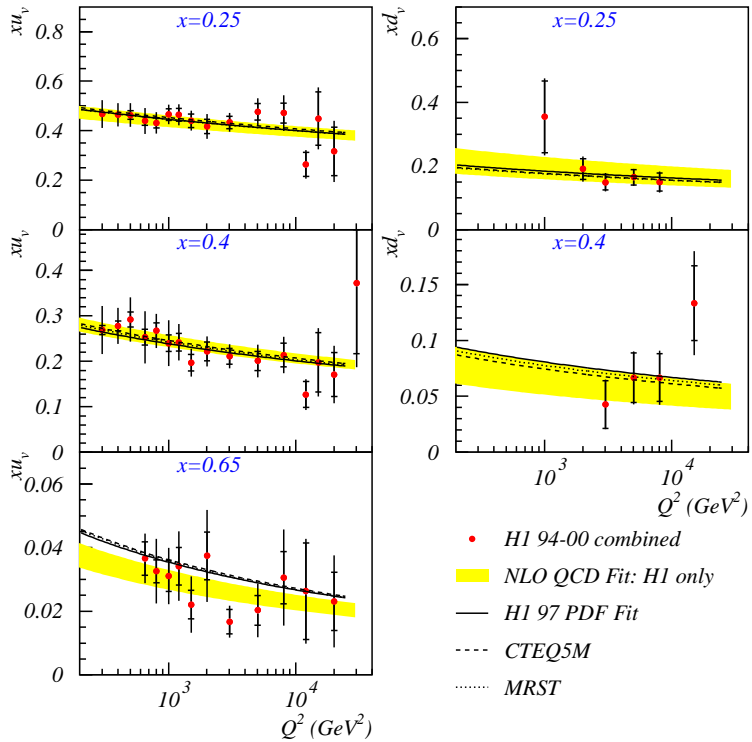,width=\textwidth}}
\end{minipage}\hspace*{-.5cm}
\begin{minipage}{0.3\textwidth}\vspace*{-4.5cm}
\caption{\label{fig=xuxd}Valence quark distributions $xu_{v}$ and 
$xd_{v}$ as determined with a NLO QCD fit (shaded error bands) using 
only cross section measurements from H1.  Data points have been 
extracted via a local subtraction method; inner (full) error bars show 
the statistical (total) error.  Other parameterizations extracted 
using also low $Q^{2}$ fixed target data are shown for comparison.}
\end{minipage}
\vspace*{-.5cm}
\end{figure}

The full sensitivity of the available $\e\p$ scattering data is 
exploited by H1 when using all $\e^{+}\p$ and $\e^{-}\p$ NC and CC 
cross sections in a combined NLO QCD fit~\cite{EPS2001:2001im}.  From 
the fit is is possible to extract the dominant valence quark 
distributions $xu_{v}$ and $xd_{v}$ at high~$Q^{2}$ and high~$x$.  The 
resulting parton densities are shown in figure~\ref{fig=xuxd} in the 
form of a band indicating the estimated uncertainty.  The result is 
compared to the MRST~\cite{Martin:1998sq} and CTEQ5~\cite{Lai:1996mg} 
parameterizations as well as the H1 97 PDF fit~\cite{Adloff:1999ah}, 
all of which include fixed target data.  The uncertainty of the parton 
densities was estimated from the experimental errors on the data 
following the prescription given in~\cite{Pascaud:1995pa}.  The 
relative precision of the $\u$-valence varies between 6\% for $x=0.25$ 
and 10\% for $x=0.65$.  The $\d$-valence is essentially constrained by 
the $\e^{+}\p$ CC data only and has a precision of about 20\%.  The 
new fit results agree with the other parameterizations within errors.  
The biggest differences appear at the highest~$x=0.65$ where the H1 
fit is about 17\% lower with little dependence in the range of $Q^{2}$ 
shown.  The difference remains within about two standard deviations.  
Using a local extraction method~\cite{Adloff:1999ah} data points are 
derived for the valence quark densities via the relation 
$xq_{v}(x,Q^{2}) = \sigma_{\rm meas}(x,Q^{2}) \left[ 
xq_{v}(x,Q^{2})/\sigma(x,Q^{2}) \right]_{\rm th}$.  Here, $\sigma_{\rm 
meas}$ is the measured NC or CC double differential cross section, 
while $[\ldots]_{\rm th}$ expresses the theoretical expectation for 
the ratio of the valence quark density and the cross section.  In 
figure~\ref{fig=xuxd} only points with an $xq_{v}$ contribution of 
greater than 70\% to the total cross section are selected.  The 
extracted parton densities are thus rather independent of the 
theoretical input as the uncertainty on the dominant valence quark 
contribution and that of the corresponding total cross section largely 
cancel in the ratio.  Compared to earlier H1 
results~\cite{Adloff:1999ah} the data provide an improved statistical 
precision for the valence quark densities by up to a factor 2; they 
are in good agreement with the global fits.

\section{Future Perspectives}
\label{sec=Future Perspectives}

In the year 2000 the HERA machine was significantly upgraded, aiming 
for a five-fold increase in luminosity~\cite{bartel}.  To this end, 
the final focus magnets for the 920 GeV proton beam are moved closer 
to the interaction regions of the two $\e\p$ collider experiments H1 and 
ZEUS. In addition spin rotators have been installed in order to 
provide polarized leptons~\footnote{The HERMES experiment uses such 
spin rotators already since 1998.}.

To adapt to this new situation and to optimally exploit the increased 
luminosity, H1 and ZEUS both improved and upgraded several detector 
components.  The modifications, in particular, concern the luminosity 
and polarization measurement, trigger capabilities, upgrade of the 
silicon vertex detectors and particle identification and 
reconstruction in the forward direction, i.e.\ the direction of the 
proton beam.  The physics return from HERA delivering $\e\p$ 
collisions with an annual integrated luminosity of $\sim$~200--250 
pb$^{-1}$ and with significant longitudinal polarization is well 
documented in the proceedings of the HERA Future Physics 
Workshop~\cite{Ingelman:1996ge}.  It comprises additional sensitivity 
to phenomena beyond the ${\cal SM}$ such as contact interactions, 
leptoquarks and flavour changing neutral currents, access to 
electroweak physics, and the capability to study rare processes in the 
regime of small momentum transfers $Q^{2}$ like deeply virtual compton 
scattering, onium production or electroproduction of open charm.  
Here, emphasis shall be put on the investigation of electroweak 
effects, in particular their utilization to disentangle the neutral 
current couplings of the $\u$- and $\d$-quark.

A major impact is expected from the provision of polarized leptons.  
Within the Standard Model NC and CC cross sections are affected not 
only by the charge of the incoming lepton beam but also by its 
longitudinal polarization; the reduced cross section for scattering 
polarized leptons off unpolarized protons is --- neglecting the 
influence of ${\cal F}_{L}$ --- given by
\begin{eqnarray}
    \label{eqn=wpol}
    \tilde{\sigma}^{\pm}_{\rm \{NC,CC\}}   & = & 
    \tilde{\sigma}^{\pm}_{0,{\rm \{NC,CC\}}} \;\pm \; {\cal P}\;
    \tilde{\sigma}^{\pm}_{\cal P,{\rm \{NC,CC\}}} \\
    \tilde{\sigma}^{\pm}_{\cal P,{\rm NC}} & = & 
    Y_{+}{\cal F}_{2}^{\cal P} \mp Y_{-} {\cal F}_{3}^{\cal P} \\
    \tilde{\sigma}^{\pm}_{\cal P,{\rm CC}} & = & 
    \tilde{\sigma}^{\pm}_{0,{\rm CC}} \;\; \raisebox{-1mm}[-1mm]{,}
\end{eqnarray}
where ${\cal P}$ denotes the lepton beam polarization, 
$\tilde{\sigma}_{0}$ represents the expression for the unpolarized 
case (${\cal P}=0$) as discussed in sections~\ref{sec=NC} 
and~\ref{sec=CC}, and $\tilde{\sigma}_{\cal P}$ contains the effects 
on the cross section due to a finite value of ${\cal P}$.  The 
definitions of ${\cal F}_{2}^{\cal P}$ and ${\cal F}_{3}^{\cal P}$ are 
similar to those in equation~\ref{eqn=F2def} with differing 
coefficient functions~\cite{Cashmore:1996je}.  As different choices of 
the lepton charge and polarization involve different contributions 
from the different terms in equation~\ref{eqn=wpol}, data sets taken 
for various charge/polarization scenarios deliver complementary 
information on couplings and parton distributions.  Neutral current 
$\e^{\pm}\p$ data taken with different beam polarizations can thus be 
used to measure electroweak couplings or to disentangle the individual 
valence quark densities, which would help to gain important 
information on the d/u ratio at high~$x$ where uncertainties are still 
large~\cite{Botje:2001fx,kuhlmann}.  Moreover, by choosing suitable 
beam configurations ${\cal SM}$ processes may be 
substantially suppressed such that the sensitivity to exotic phenomena 
is enhanced significantly.

\begin{figure}[tb]
\begin{minipage}{0.496\textwidth}
\centerline{\epsfig{figure=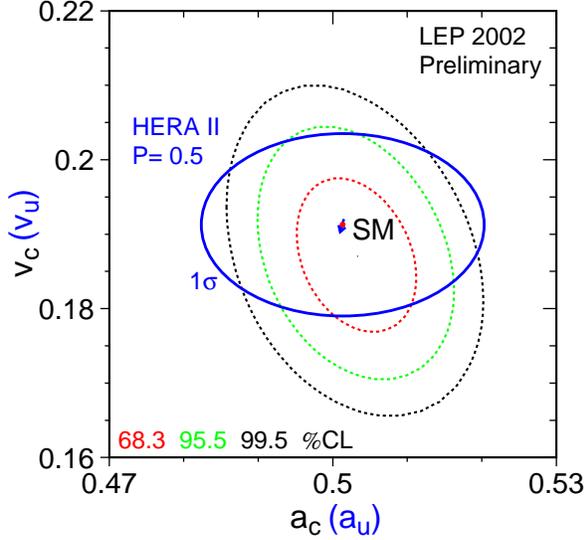,width=\textwidth}}
\end{minipage}
\begin{minipage}{0.446\textwidth}\vspace*{2cm}
\caption{\label{fig=cont}Comparison of the most recent LEP 
measurement~\protect\cite{myatt} of the $Z^{0}$ vector and axial 
couplings to the $c$-quark (dashed lines showing the $1\sigma$-, 
$2\sigma$- and $3\sigma$-contour) with a potential HERA result (full 
ellipse) for the corresponding $u$-quark couplings based on 
1000~pb$^{-1}$.}
\end{minipage}
\vspace*{-.5cm}
\end{figure}

With 250~pb$^{-1}$ of NC data for each lepton charge and a 
polarization of ${\cal P}=\pm 70$\% it has been estimated that the 
$\Z^{0}$ vector coupling $v_{u}$ ($v_{d}$) and the axial vector 
coupling $a_{u}$ ($a_{d}$) of the $\u$-quark ($\d$-quark) can be 
measured with accuracies of 13\% (17\%) and 6\% (17\%), 
respectively~\cite{Cashmore:1996je}.  Figure~\ref{fig=cont} shows the 
expected accuracy of $v_{u}$ vs.\ $a_{u}$ for a more realistic value 
of ${\cal P}=\pm 50$\% in comparison to the most recent LEP results on 
$c$-quark couplings~\cite{myatt}.  Both measurements are complementary 
as they probe different quark flavours.  The HERA measurement would 
thus provide a stringent test of the consistency of the Standard 
Model.

\section{Conclusion}

Recent HERA measurements of $\e^{-}\p$ and $\e^{+}\p$ scattering cross 
sections at high momentum transfers $Q^{2}$ have been presented.  The 
data are in good agreement with the Standard Model predictions and 
show a clear separation between the $\e^{-}\p$ and $\e^{+}\p$ cross 
sections in agreement with the expectation from electroweak theory of 
a positive (negative) contribution from $\gamma\Z$ interference to 
$\e^{-}\p$ ($\e^{+}\p$) scattering.  
Utilization of the observed differences reveals information on the 
valence quark densities through extraction of the structure functions 
$x{\cal F}_{3}$ and $xF_{3}^{\gamma\Z}$ and a combined fit to 
NC and CC $\e\p$ scattering data.  Future data, to be taken with the 
upgraded HERA machine during the coming years, will give additional 
insight into the valence quark content of the proton and allow a 
precise measurement of the electroweak $\Z$ vector and axial couplings 
of the light quarks.

\section*{Acknowledgments}
This work was supported by the BMBF (contract no.  05-H1-1PEA/6) and 
the European Union.


\section*{References}
\begin{thebibliography}{99}
%
\bibitem{Adloff:1999ah}
C.~Adloff {\it et al.}  [H1 Collaboration],
Eur.\ Phys.\ J.\ C {\bf 13}, (2000) 609.
\bibitem{Adloff:2000qj}
C.~Adloff {\it et al.}  [H1 Collaboration],
Eur.\ Phys.\ J.\ C {\bf 19}, (2001) 269.
%
\bibitem{ICHEP2000:2000cm}
ZEUS Collaboration, 
``Measurement of high-$Q^{2}$ Charged Current Cross Section in $\e^{-}\p$ Deep Inelastic Scattering at HERA'',
EPS 01, Budapest, 2001.  
%
\bibitem{EPS2001:2001im}
H1 Collaboration, 
``Inclusive Measurement of Deep Inelastic Scattering at high $Q^{2}$ in $\e\p$ Collisions at HERA'', 
EPS 01, Budapest, 2001.  
%
\bibitem{EPS2001:2001np}
ZEUS Collaboration, 
``Measurement of high-$Q^{2}$ Neutral Current Cross Sections in $\e^{+}\p$ Deep Inelastic Scattering at HERA'', 
EPS 01, Budapest, 2001.  
%
\bibitem{EPS2001:2001cp}
ZEUS Collaboration, 
``Measurement of high-$Q^{2}$ Charged Current Cross Sections in $\e^{+}\p$ Deep Inelastic Scattering at HERA'',
EPS 01, Budapest, 2001. 
%
\bibitem{EPS2001:2001nm}
ZEUS Collaboration, 
``Measurement of high-$Q^{2}$ Neutral Current Cross Sections in $\e^{-}\p$ DIS and Extraction of the Str.\ Function $xF_{3}$ at HERA'', 
EPS 01, Budapest, 2001. 
%
\bibitem{Lai:1999wy}
H.~L.~Lai {\it et al.}  [CTEQ Collaboration],
Eur.\ Phys.\ J.\ C {\bf 12} (2000) 375.
%
\bibitem{Derman:sp}
E.~Derman,
Phys.\ Rev.\ D {\bf 7}, (1973) 2755.
%
\bibitem{Ingelman:1987zv}
G.~Ingelman and R.~R\"uckl,
Phys.\ Lett.\ B {\bf 201}, (1988) 369.
%
\bibitem{Rizvi:2000qf}
E.~Rizvi and T.~Sloan,
arXiv:hep-ex/0101007.
%
\bibitem{Gross:1969jf}
D.~J.~Gross and C.~H.~Llewellyn Smith,
Nucl.\ Phys.\ B {\bf 14}, (1969) 337.
%
\bibitem{Martin:1998sq}
A.~D.~Martin {\it et al.},
Eur.\ Phys.\ J.\ C {\bf 4} (1998) 463.
%
\bibitem{Lai:1996mg}
see e.g.\ H.~L.~Lai {\it et al.},
Phys.\ Rev.\ D {\bf 55}, (1997) 1280.
%
\bibitem{Pascaud:1995pa}
C.~Pascaud and F.~Zomer, LAL preprint, LAL/95-05, 1995.
%
\bibitem{bartel}
W.~Bartel {\it et al.},
Workshop on Future Physics at HERA, Hamburg, 1996.
%
\bibitem{Ingelman:1996ge}
G.~Ingelman, A.~De Roeck and R.~Klanner,
``Future physics at HERA'', DESY-96-235.
%
\bibitem{Cashmore:1996je}
R.~J.~Cashmore {\it et al.},
Workshop on Future Physics at HERA, Hamburg, 1996.
%
\bibitem{Botje:2001fx}
M.~Botje,
J.\ Phys.\ G {\bf 28} (2002) 779.
\bibitem{kuhlmann}
S.~Kuhlmann {\it et al.},
Phys.\ Lett.\ B {\bf 476} (2000) 291.
\bibitem{myatt}
G.~Myatt, these proceedings.

\end{thebibliography}
\end{document}